\documentclass[preprint,aps,prd]{revtex4}
\usepackage{epsfig}
\newcommand \be {\begin{equation}}
\newcommand \bea {\begin{eqnarray}}
\newcommand \ee {\end{equation}}
\newcommand \eea {\end{eqnarray}}
\newcommand \bed {\begin{displaymath}}
\newcommand \eed {\end{displaymath}}

\newcommand{\bit}{\begin{itemize}}
\newcommand{\eit}{\end{itemize}}

\begin{document}
\topskip 2cm

\begin{titlepage}

\rightline{\today}

\begin{center}
{\Large \bf NONUNIFORM SYMMETRY BREAKING IN NONCOMMUTATIVE $\lambda \Phi^4$ THEORY} \\
\vspace{2.5cm}
{\large Paolo Castorina$^{1,2}$, Dario Zappal\`a$^{2,1}$} \\
\vspace{.5cm}
{\sl $^{1}$ Department of Physics, University of Catania} \\
{\sl via S. Sofia 64, I-95123, Catania, Italy}\\
\vspace{.3cm}
{\sl $^{2}$ INFN, Sezione di Catania, via S. Sofia 64, I-95123, Catania, Italy } \\
\vspace {.6cm}
{\sl e-mail: paolo.castorina@ct.infn.it, dario.zappala@ct.infn.it}

\vspace{1.5cm}

\begin{abstract}
The spontaneous symmetry breaking in noncommutative $\lambda\Phi^4$ theory has been analyzed by using the formalism
of the effective action for composite operators in the Hartree-Fock approximation. It turns out that there is no phase transition
to a constant vacuum expectation of the field and the broken phase corresponds to a nonuniform background. 
By considering $<\phi(x)>=A~\cos(\vec Q \cdot \vec x)$ the generated mass gap depends on the angles among the momenta 
$\vec k$ and $\vec Q$ and the noncommutativity parameter $\vec\theta$. The order of the transition is not easily determinable
in our approximation. 
\end{abstract}

\end{center}

\vspace{2 cm}


INFNCT/01/03 - March 2003


\end{titlepage}

\section{ INTRODUCTION}

\par 
The  effects of  noncommuting  coordinates have recently received new attention in relation  with string theories \cite{connes,witten}
and there is a  great effort in understanding the fundamental properties of noncommutative field theories.
In particular, the phase structure of   $\lambda \Phi^4$ theory has been recently discussed \cite{camp,gubser,chen,rivelles} 
(see also \cite{bieten} and \cite{catterall} for numerical studies of the theory in three and two Euclidean dimensions)
and  Gubser and Sondhi \cite{gubser} showed that  there are indications for  a first order phase 
transition to a nonuniform ground state due to noncommutativity. 
\par In this paper  we essentially address the problem of spontaneous symmetry breaking  within the formalism
of the Effective Action for composite operators introduced by Cornwall, Jackiw and Tomboulis \cite{cjt} (CJT), 
in the Hartree-Fock (HF) approximation. In this approach  we have coupled  
extremum equations, for the field and the full propagator, which shed new light on the transition from the ordered to  the disordered  phase.
\par We work in the cutoff field theory mainly for two reasons. First of all it is  
 not yet  clear whether the noncommutative theory
is renormalizable \cite{minwalla,mvan,chepelev,griguolo,doplicher,ruiz} and
moreover the renormalization of the Effective Potential
in the HF approximation is cumbersome also for the commutative case \cite{pi1,noi} .
Nevertheless the proposed approach gives interesting indications on the phase of the theory.
In particular we find in the HF approximation  that:

a) The transition from $< \phi > =0$ to $< \phi > \neq 0$ turns out     
first order also for the commutative theory.

b) For the noncommutative theory, the minimization of the Effective Action has 
no solution for  $< \phi > = constant \ne 0 $ and the broken phase   corresponds to a  nonuniform background field.

c) in the nonuniform stripe phase, with $< \phi(x) > = A \cos( Q \cdot x)$ \cite{gubser,braz}, the generated  mass gap
depends on  the value of  $k_\mu \theta_{\mu \nu} Q_\nu$
where $k$ is the momentum,  and
\begin{equation}
\label{uno}
[x_{\mu},x_{\nu}]= i \theta _{\mu \nu} 
\end{equation}
are  the coordinate commutators.

The  paper is organized as follows. In Section II we briefly review the CJT formalism and apply it in the HF approximation to the 
commutative $\lambda \Phi^4$ theory; Section  III is devoted to the noncommutative case with $\phi = costant$;
 in Section IV we study the stripe phase and Section V contains the conclusions.
 
\section{ COMMUTATIVE $\lambda \Phi^4$ THEORY } 
 
 In this Section we shall briefly summarize the Effective Action 
for composite operator as introduced by Cornwall, Jackiw and Tomboulis (CJT) 
(see \cite{cjt} for details)  and study the spontaneous symmetry breaking in the commutative case.
The CJT Effective Action $\Gamma (\phi , G)$ is  given by
\begin{equation}
\label{due}
\Gamma (\phi , G) ={ I(\phi) +{ i \over 2} Tr Ln {D G^{-1}} +{i \over 2} Tr {\Delta ^{-1}(\phi) G} + \Gamma_{2}(\phi, G)-{i \over 2} Tr1}
\end{equation}
where $\phi(x)$ is the expectation value of the field on the ground state, $ G(x,y)$ is the full
connected propagator of the theory, $I(\Phi)$ is 
 the classical Effective Action:

\begin{equation}
\label{tre}
 I(\Phi) = \int d^4 x~L(x)
\end{equation}
$D$ is the free propagator

\begin{equation}
\label{quattro}
i D^{-1} (x-y)= -{(\partial^{\mu} \partial_{\mu} + m ^2)}{ \delta^4 (x-y)}
\end{equation}
and 

\begin{equation}
\label{cinque}
i \Delta ^{-1} (x-y)= -{(\partial^{\mu} \partial_{\mu} + m^2)} { \delta ^4(x-y)} +
 {{\delta ^2 I_{int}(\phi)} \over {\delta \phi (x) \delta \phi (y)}}
\end{equation}
with the interaction terms $I_{int}(\phi)$  at least cubic in the fields.

The term $ \Gamma _2 (\phi, G)$  is computed as follows. In the classical action $I(\Phi)$ shift 
the field $\Phi$ by $ \phi (x)$.
The new action $I(\Phi + \phi)$ possesses terms cubic and higher 
in $\Phi$ which define an ``interaction'' part $I_{int} (\phi, \Phi)$ where
the vertices depend on $\phi(x)$.
$ \Gamma_{2}(\phi, G)$ is given by  all two particle irreducible vacuum graphs 
in the theory with vertices determined by  
$I_{int} (\phi, \Phi)$ and propagator set equal to $G(x,y)$.
The usual Effective Action is recovered by extremizing $\Gamma (\phi , G)$with respect to $G$.
\par We evaluate $\Gamma (\phi , G)$  for  the commutative  $\lambda \Phi^4$ theory with action

\begin{equation}
\label{sei}
I(\phi)=\int d^4 x \left (  {1 \over 2} 
\partial_{\mu} \phi ~ \partial^{\mu} \phi
 - {1 \over 2} m^2 \phi ^2 - { \lambda 
\over 4!} \phi ^4\right )
\end{equation}
in the Hartree-Fock approximation which corresponds to retain only the 
lowest order contribution in coupling constant to $ \Gamma_{2}(\phi, G)$ (see \cite{cjt})).
\par The coupled equations for the extrema of  $\Gamma (\phi , G)$ are

\begin{equation}
\label{sette}
{{\delta \Gamma (\phi , G)} \over {\delta \phi}} = 0
~~~~~~~~~~~~~~~~~~~~~~~~~~~~~~~~~~~~~~~~~~
{{\delta \Gamma (\phi , G)} \over {\delta G}} = 0
\end{equation}

It turns out that in this approximation the propagator can be conveniently  
parametrized as \cite{cjt,pi1,noi}
 
\begin{equation}
\label{otto}
 G(x,y)=i \int \frac{d^4 p}{(2\pi)^4} {{e^{-ip(x-y)}} \over {p^2 - M^2(p^2)}}
\end{equation}

and the two previous equations become

\begin{equation}
\label{dieci}
0= (\partial_{\mu} \partial^{\mu} + m^2) {\phi (x)} + {\lambda \over 6} \phi^3 (x) + {\lambda \over 2} \phi (x) G(x,x)
\end{equation}

\begin{equation}
\label{undici}
M^2=  m^2 + {\lambda \over 2} \phi^2  + {\lambda \over 2} G(x,x)
\end{equation}

The extrema of the Effective Action are for $\phi$ and $M$  constant. 
The previous equations contain divergences that are regularized by introducing a  cutoff $\Lambda$.
By requiring that  physical quantities are exponentially decoupled from the cuf-off we redefine the parameter $m^2$ 
to cancel the quadratic divergences,
i.e. 

\begin{equation}
\label{dodici}
\mu ^2=  m^2   + {\lambda \over 2} \int \frac{d^4 p}{(2\pi)^4} {{i} \over {p^2}}
\end{equation}
and  the coupled equations (in the  Euclidean space) become

\begin{equation}
\label{tredici}
M^2=  \mu ^2 + {\lambda \over 2} \phi ^2  + {\lambda \over 32 \pi ^ 2} M^2 \ln {M^2}
\end{equation}

\begin{equation}
\label{quattordici}
0= \phi \left ( {\lambda \over 3} \phi^2  -  M^2\right ) 
\end{equation}
where all the dimensional quantities have been rescaled in units of the cutoff $\Lambda$.
The extremum equations have two sets of solutions:

\begin{equation}
\label{quindici}
\phi = 0 
\end{equation}

\begin{equation}
\label{sedici}
M^2=  \mu ^2  + {\lambda \over {32 \pi ^ 2}} M^2 \ln{M^2}
\end{equation}
and 

\begin{equation}
\label{diciassette}
\phi ^2 = {{3 M^2} \over  \lambda}
\end{equation}

\begin{equation}
\label{diciotto}  
{1 \over 2} M^2  +   \mu ^2 = - {\lambda \over {32 \pi ^ 2}} M^2 \ln {M^2}
\end{equation}

\par Now let us consider the second set of solutions  
which is the relevant one for the spontaneous symmetry breaking,
and solve it for various values of $ \mu ^2$.

\vskip 20 pt

\par a) $ \mu ^2$ =0 

\noindent
In this case, beside the solution $M^2 =0$, $\phi =0$ obtained form Eqs. (\ref{quindici},\ref{sedici}) \cite{nota}, one finds 
two nonvanishing opposite  solutions for  $\phi$:

\begin{equation}
\label{diciannove}
\phi  = \pm{\sqrt{ {{3 M_o^2} \over \lambda}}} 
\end{equation}
with 

\begin{equation}
\label{venti}
 M_o^2 =  e^{ - { {(16 \pi ^ 2)}  \over \lambda}} 
\end{equation}

\par b) $ \mu ^2 \neq 0$  

\noindent 
To solve Eq. (\ref{diciotto}) one rescales all quantities  in unit of $M_o^2$ , the solution for 
$ \mu ^2$ =0, thus obtaining 

\begin{equation}\label{ventuno}
\hat  \mu ^2 =  - {\lambda \over 32 \pi ^ 2} \hat  M^2 \ln { \hat M^2}
\end{equation}
where $\hat  \mu ^2 = {\mu^2 / {M_o^2}}$ and  $ \hat  M^2 = {M^2 / {M_o^2}}$.
It is easy to verify that for $\hat \mu ^2 > \lambda / ( e 32 \pi ^2)$ there is no solution of 
Eq. (\ref{ventuno}) 
and the only solution of Eqs. (\ref{quindici}-\ref{diciotto}) is $\phi =0$ with a nonvanishing  mass
$M^2$ obtained from Eq. (\ref{sedici}).
In the region $0 < \hat \mu ^2 <  \lambda / ( e 32 \pi ^2)$  
there are two  solutions of 
Eq. (\ref{ventuno}) , $\hat M_1^2$ and  $\hat M_2^2$,
and then  five extrema corresponding to $\phi =0$ and to

\begin{equation}\label{ventidue}
{\hat \phi _{1,2} = \pm { \sqrt  { {3 \over \lambda } \hat M _{1,2}^2}}}
\end{equation}
Note that for $\hat \mu ^2 =  \lambda / ( e 32 \pi ^2)$ 
the two solutions for $\hat M^2$ coincide: $\hat M_1^2=\hat M_2^2$.
and correspondingly there are three different extrema in  $\phi$. 
For $ \hat  \mu ^2 < 0$ there is only one solution in $ \hat  M^2$ 
which corresponds to two nonvanishing extrema in $\phi$.
\par 
Let us finally translate the previous informations on the shape of the  Effective Potential
as a function of $ \phi$ for different values of $\mu^2$. 
For $\hat \mu ^2 > \lambda / ( e 32 \pi ^2)$ there is only the extremum at $\phi =0$ and 
the potential corresponds to plot (a) in Figure 1.
For $\hat \mu ^2 =  \lambda / ( e 32 \pi ^2)$ two new nonvanishing extrema  appear and 
for  $0 < \hat \mu ^2 <  \lambda / ( e 32 \pi ^2)$  there are five extrema and the shape is 
as in plot (b). Note that when lowering 
$\hat \mu ^2$, the maxima of the potential for $\phi \neq 0$ decrease and the corresponding 
values of $\phi$ become smaller and approach zero and also  the minima decrease but 
the corresponding values of  $\phi$ increase.  
For $ \hat \mu ^2$ =0 the solution corresponding to the maxima have 
merged into $\phi = 0$ and there are three extrema (see plot (d)).
Then for some critical, finite and positive value of $ \hat \mu ^2$ the  potential 
must be of the form reported in plot (c),
with three degenerate minima at different values of $\phi$.
This picture implies a (weak) first order phase transition and suggest that in this case 
the HF approximation  gives  a ``coarse grain'' description reliable to establish the 
occurrence of the transition but probably not its order\cite{polch}.

\section{  NONCOMMUTATIVE  $\lambda \Phi^4$ THEORY}

In this Section we shall analyze the extremum equations for the CJT  Effective Action 
for the noncommutative theory defined by the action
 
\begin{equation}
\label{ventitre}
 I(\phi)=\int d^4 x \left (  {1 \over 2} 
\partial_{\mu} \phi ~ \partial^{\mu} \phi
 - {1 \over 2} m^2 \phi ^2 -{  \lambda 
\over 4!} \phi ^{4*}\right )
\end{equation}
 where the star product is defined by ($i,j =1,.,4$)

\bea
&&\phi ^{4*}(x)= \phi(x) * \phi(x) * \phi(x) * \phi(x) = \nonumber\\
\nonumber\\
\label{ventiquattro}
&&\exp\left \lbrack{ i \over 2} \sum_{i<j}\theta_{\mu \nu} \partial^{\mu}_{x_i} \partial^{\nu}_{x_j}\right \rbrack
\Bigl ( \phi (x_1) \phi (x_2) \phi (x_3) \phi (x_4)\Bigr ) \left |_{x_{i}=x} \right.
\eea

The theory has been discussed in the literature (see e.g. the review \cite{douglas}) and  
the planar approximation, $\theta \Lambda^2\to \infty$ has the same behavior of the commutative theory: 
a phase transition for $<\phi>=\phi_0=const$ and a translational invariant full propagator 
parametrized as in Eq.(\ref{otto}) with constant $M$ .

Let us now check whether this behavior survives to the genuine noncommutative effects, i.e. for finite 
$\theta \Lambda^2$. With a translational invariant propagator
\be
G(x,y)=\int \frac {d^4 p}{(2\pi)^4} e^{-ip(x-y)} G(p)  
\ee
the CJT Effective Action in momentum space reads

\bea
 &&\Gamma (\phi , G) = {1 \over 2} \int { d^4 p \over {(2 \pi)^4}} (p^2 - m^2) \phi (p) \phi (-p)
\nonumber \\
&& - {\lambda \over {4!}} \left [ {\prod _{a=1}^{4} \int { d^4 p_a \over {(2 \pi)^4}} \phi (p_a)}
\right ] \delta ^4 
\left (\sum_a{p_a}\right) \exp\left ({i \over 2} p_1 \wedge p_2\right)
\exp\left ({i \over 2} p_3 \wedge p_4\right) \nonumber \\ 
&& + {i \over 2} \delta^4 (0) \int { d^4 p \over (2 \pi)^4} \ln D(p)G^{-1}(p) + 
 {1 \over 2} \delta^4 (0) \int { d^4 p \over (2 \pi)^4} (p^2 - m^2) G(p) \nonumber \\
&& -  { \lambda \over 6} \int { d^4 p \over (2 \pi)^4} 
\int { d^4 q \over (2 \pi)^4}
 \phi(p) \phi(-p) G(q) \left [1 + {1 \over 2} \exp {(i  q \wedge p)}\right ] \nonumber \\
\label{venticinque}  
&& -   { \lambda \over {12}} \delta^4 (0) 
\int { d^4 p \over (2 \pi)^4} \int { d^4 q \over (2 \pi)^4} G(p)G(q)
 \left [1 + {{1 \over 2}} \exp {(i  q \wedge p)}\right ] 
\eea
\noindent 
where $q \wedge p\equiv q_\mu \theta^{\mu \nu} p_\nu$.
In the noncommutative case let us parametrize 
\be 
G(q)=\frac{i}{q^2-M^2(q)}
\ee
where $M^2$ is a function of the four-momentum.

From Eq. (\ref{venticinque}) we get two coupled extemum equations for $M^2(q)$ and $\phi(q)$
\bea
\delta^4 (0) &&\left  \lbrack M^2(q) -  m^2 - {\lambda \over 3} \int \frac {d^4 p}{(2\pi)^4}
~ {  i  \over { (p^2 - M^2(p))}} \left (1 + {1\over 2}  \exp (i  q \wedge p) \right ) \right \rbrack
\nonumber\\
\label{mextr}
&&=
{\lambda \over 3} \int \frac {d^4 p}{(2\pi)^4}\phi(p)\phi(-p) \left (1 + {1\over 2}  \exp (i  q \wedge p) \right )
\eea

\bea
&&\left  \lbrack q^2 -  m^2 - {\lambda \over 3} \int \frac {d^4 p}{(2\pi)^4}
~ {  i  \over { (p^2 - M^2(p))}} \left (1 + {1\over 2}  \exp (i  q \wedge p) \right ) \right \rbrack \phi(-q) \nonumber\\
&&={\lambda \over 12} \int \frac {d^4 p}{(2\pi)^4}\int \frac {d^4 k}{(2\pi)^4}
\phi(p)\phi(k-p)\phi(-q-k)  \exp (\frac{i}{2}  p \wedge k)  \nonumber\\
\label{fiextr}
&&\times \left ( \exp (\frac{i}{2}  k \wedge q)+ \exp (-\frac{i}{2}  k \wedge q)
 \right )
\eea

With the help of Eq. (\ref{mextr}), one can get rid of the constant $m^2$ in Eq. (\ref{fiextr}) which  becomes

\bea
&&\left  \lbrack q^2 - M^2(q)  \right \rbrack \phi(-q)
 +(\delta^4 (0))^{-1} \phi(-q)  {\lambda \over 3} \int \frac {d^4 p}{(2\pi)^4}
\phi(p)\phi(-p) \left (1 + {1\over 2}  \exp (i  q \wedge p) \right )\nonumber\\
&&={\lambda \over 12} \int \frac {d^4 p}{(2\pi)^4}\int \frac {d^4 k}{(2\pi)^4}
\phi(p)\phi(k-p)\phi(-q-k)  \exp (\frac{i}{2}  p \wedge k) \nonumber\\
\label{fiextr2}
&& \times \left ( \exp (\frac{i}{2}  k \wedge q)+ \exp (-\frac{i}{2}  k \wedge q)
 \right )
\eea
Then, analogously to what has been done for the commutative case 
in Eq. (\ref{dodici}), we cancel the quadratic divergence in Eq. (\ref{mextr}) by defining 
\begin{equation}\label{ventotto}
\mu ^2=  m^2   + {\lambda \over 3} \int \frac{d^4 p}{(2\pi)^4} ~{{i} \over {p^2}}
\end{equation}
and we can directly check whether a constant background
\begin{equation}\label{cam}
\phi(q)=\phi_o \delta^4(q)
\end{equation}
is a solution of the extremum equations. Indeed, by replacing Eqs. (\ref{ventotto}) and (\ref{cam})  in 
 Eqs. (\ref{mextr}) and (\ref{fiextr2}) we get 
\begin{equation}
\label{ventinove}
M^2(q)= \mu ^2 + {\lambda \over 2} \phi _o^2    + {\lambda \over 3}\int \frac {d^4 p}{(2\pi)^4} ~ {i M^2(p) 
\over {{p^2} (p^2 - M^2(p))}}
+ {\lambda \over 6} \int \frac {d^4 p}{(2\pi)^4} ~ {  i  \exp (i  q \wedge p)  \over { (p^2 - M^2(p))}} 
\end{equation}

\begin{equation}
\label{trenta}
0= \phi_o \left ( {\lambda \over 3} \phi_o^2  -  M^2(q)\right ) \delta^4 (q) 
\end{equation}

We note that in Eq.  (\ref{trenta}) we can replace $M^2(q)$ with $M^2(0)$ because of the delta function $ \delta^4 (q)$.
As usual there is the solution   $\phi _o = 0$ and $M^2(q)$ given by Eq. (\ref{ventinove}) (where the term proportional 
to the field $\phi_0$ has been discarded).  

This case has been studied in \cite{gubser} with the interesting result that for  $q^2 \rightarrow 0$ the function $M^2(q)$ has a 
singular behavior
($c$ constant)

\begin{equation}\label{trentadue}
M^2(q) \rightarrow \frac {c}{(q^2/\Lambda^2)(\theta\Lambda^2)} 
\end{equation}
which  is  a genuine effect of  the noncommutative structure of the theory and does  not change if one considers the same equation
for $\phi _o \ne 0$ because it is due to the phase factor in the integral in Eq. (\ref{ventinove}).

Let us consider the equations (\ref{ventinove}) and  (\ref{trenta})
in the case of constant finite and nonvanishing background $\phi _o \ne 0$. 
As noticed above, due to the noncommutative terms, $M^2$  constant is not a solution 
of Eq.(\ref{ventinove}):
$M^2(q)$ must depend on $q$ and moreover $M^2(q)$ for small $q$ is singular as in Eq. (\ref{trentadue}).
Then the condition  $ (\lambda / 3) \phi_o^2  -  M^2(0)=0 $ from Eq. (\ref{trenta}) does not admit a 
finite constant solution $\phi_o$.
Therefore a finite  constant solution  $\phi_o \ne 0$ is ruled out by the analysis of the combined 
equations.

It is interesting to note  that an indication of the impossibility of finding a constant field solution of our extremum
equations could have been obtained directly from Eqs. (\ref{mextr}) and  (\ref{fiextr}). In fact after  replacing in these 
equations the constant field solution Eq. (\ref{cam}), the quadratic divergences that
appear in  the two extremum equations cannot be simultaneously cancelled by a single counterterm, 
namely $m^2$ as fixed in Eq. (\ref{ventotto}), and therefore all solutions of the coupled equations
are plagued with a quadratically divergent integral. 

In our previous analysis the problem of cancelling the quadratic divergences 
has been hidden by the replacement performed to get Eq.  (\ref{fiextr2})
which apparently made the case of constant background field free of divergences
although in the end  we could not find any suitable solution because of
the singular behavior of $M^2(q)$  at $q=0$ shown in Eq. (\ref{trentadue}).
By looking at  Eq. (\ref{mextr}), it is easy to realize that this $q$ dependent singular behavior 
is directly related to the not complete cancellation of the quadratic divergences for 
finite $\theta\Lambda^2$. Indeed this pathology is not present in the planar limit. 

In conclusion, we have to reject the constant solution  $\phi_o \ne 0$
and look for  spontaneous symmetry breaking  
only  in  a nonuniform phase.

\section  { THE STRIPE PHASE}  

\par As pointed out in \cite{braz} the phase transition to a nonuniform state is related to a periodic correlation function 
$<\phi(x) \phi(0)>$ which oscillates
 in sign for large $x$. For this reason we consider 
a time independent stripe pattern 
\begin{equation}
\label{trentatre}
<\phi(\vec x)> =A \cos (\vec Q \cdot \vec x)
\end{equation}
and calculate the CJT Effective Action in the Hartree-Fock approximation in the static limit \cite{cjt}.
Let us then assume that $\theta^{\mu\nu}$ has no time component $\theta^{0 i}=0$ and  $\theta^{ij}=\varepsilon^{ijk}\theta_k$.

It is impossible to study the transition to the stripe phase with the most general
class of propagators $G$ and we shall limit ourselves to a Raileigh-Ritz variational approach where, however
a meaningful  ansatz for $G$ requires at least some physical indications on  its  asymptotic behaviors.
Indeed the nonuniform background given in Eq. (\ref{trentatre}) has a new typical scale $|\vec x|\sim 1/|\vec Q|$. 
For small $|\vec Q|$ (in cutoff units), the effect of the nonuniform
background will be relevant only for large distances and the background will be a slowly varying function of $\vec x$.

Then for momenta $|\vec p| >>|\vec Q|$, the breaking of the translational and rotational invariance is expected negligible 
and a good ansatz for the tridimensional propagator in momentum space is
\be 
G(\vec p)=\frac{1}{2 \sqrt{\vec p^2+ M^2_o}}
\ee
where, analogously to the constant background case, $M^2_o$ is a constant.

In the region $|\vec p| < |\vec Q|$, the previous ansatz is of course not reliable and to obtain 
further informations on the behavior of $G$ let us preliminarily assume that  the breaking of the
translational invariance appears in the field expectation value only, i.e. in Eq. (\ref{trentatre}), 
while we consider a general translational invariant form of the propagator with  
\be
\label{gi} 
G(\vec p)=\frac{1}{2 \sqrt{\vec p^2+ M^2(\vec p)}}
\ee

Then we analyze the extremum equations obtained by minimizing, with respect to $M^2(\vec p), ~A$ and $Q^2$, the quantity $E(\phi,G)$
defined as
\be
\label{ene} 
- \delta(0) E(\phi,G)=\Gamma(\phi,G)
\ee
with $ \Gamma(\phi,G)$ computed in the static limit \cite{cjt}.
The three coupled equation for $M^2(\vec p), ~A$ and $Q^2$ respectively turn out
   
\bea
&&M^2(\vec p) = m^2
+ A^2 {\lambda \over 6}\left (1 + {1 \over 2} \cos {(  \vec p  \wedge \vec Q)}\right)
\nonumber \\
\label{trentaquattro}
&&+ {\lambda \over 3} \int \frac{d^3 k}{(2\pi)^3} \frac{1}{2\sqrt{\vec k^2+M^2(\vec k)}}
\left(1 + {1 \over 2} \cos {(  \vec p  \wedge \vec k)}\right)
\eea

\begin{equation}
\label{trentacinque}
A\left \lbrack  Q^2 +m^2 + A^2 {\lambda \over 8}
+ {\lambda \over 3} \int \frac {d^3 k}{(2\pi)^3} \frac{1}{2\sqrt{\vec k^2+M^2(\vec k)}}\left (1 + {1 \over 2} \cos {(  \vec Q  \wedge \vec k)}\right)
\right\rbrack=0
\end{equation}

\begin{equation}\label{trentasei}
Q^2  -  {\lambda \over 12} \int \frac{d^3 k}{(2\pi)^3} \frac{1}{2\sqrt{\vec k^2+M^2(\vec k)}}( \vec k  \wedge \vec Q)\sin {( \vec k  \wedge \vec Q)}
=0
\end{equation}
 
The cancellation of the quadratic divergences is now obtained by defining 

\begin{equation}\label{newmu}
\mu ^2=  m^2   + {\lambda \over 3} \int \frac{d^3 k}{(2\pi)^3} \frac{1}{2 |\vec k |}
\end{equation}

Let us first discuss Eq. (\ref{trentasei}) and look for a small $Q^2$ solution.
Due to the strong oscillating factor, for small $Q$ the integration region is dominated by large $k$
and then we can replace $G$ with its asymptotic behavior in Eq. (\ref{gi}) or, in other words, 
$M^2(\vec k)\sim M^2_o$.

By choosing the configuration $\vec \theta=(0,0,\theta)$ and $\vec  Q=(Q/\sqrt{2},Q/\sqrt{2},0)$, 
the small $Q^2$  selfconsistent solution turns out 

\be\label{qqad}
\frac{Q^2}{\Lambda^2}=\left ( {\lambda \over {24 \pi^2 }}\right )^{1/2} \frac{1}{\theta\Lambda^2}
\ee
where we consider from now on large but finite values of $\theta\Lambda^2$.

The next step is to consider the gap equation (\ref{trentaquattro}).
As previously discussed for large $p^2>>Q^2$ one expects $M^2(\vec p) \sim M^2_o$.
For  $p^2<<Q^2$ the selfconsistent behavior of $M^2(\vec p)$ is

\be\label{masy}
M^2(\vec p)\left. \right|_{p\to 0} \sim \alpha +{ {A^2 \lambda} \over {12}}\left \lbrack 1 + {1\over 2}\cos \left (\vec p \cdot ( \vec Q  \times \vec \theta)
\right ) \right \rbrack
+{ {\lambda} \over {6\pi^2 }}\frac{1}{|\vec p  \times \vec \theta|^2}
\ee
where $\alpha$ is a constant and  $\times$ indicates the usual vector product. 
The last term corresponds to the small $p$ contribution analogous to the 
singular behavior in Eq. (\ref{trentadue}).

Finally one can qualitatively analyze the phase transition by looking at the equation for $A$, Eq.  (\ref{trentacinque}),
that  can be written in the form 
\bea
&&A^2=-\frac{8}{\lambda} \Biggl \lbrace  Q^2 +\mu^2      \nonumber\\
\label{aquad}
&& +\frac{\lambda}{3} \int \frac {d^3 k}{(2\pi)^3}   \frac{1}{2\sqrt{\vec  k^2 +M^2(\vec k)}}
\left (1-\frac {\sqrt{ \vec k^2 +M^2(\vec k)}}{| \vec k |}+{1\over 2} \cos\left (\vec k \cdot ( \vec Q  \times \vec \theta)\right )\right ) \Biggr \rbrace
\eea

Indeed, if the contribution of the integrals in Eq. (\ref{aquad}) is negligible, a solution $A^2\neq 0$ for $Q^2\neq 0$
is allowed only for sufficiently negative $\mu^2$. 
Moreover in this approximation the energy gap between the stripe phase and the $A=0$ phase is about 
$\Delta E\sim -\lambda A^2/64$, showing that the broken phase is energetically favored.

On the basis of this preliminary analysis, to establish the occurrence of the spontaneous symmetry breaking
one can apply the following settings of selfconsistent approximations for the numerical 
evaluation of the energy defined in Eq. (\ref{ene}):

1) slowly changing background, i.e.
\be
\label{qcond}
\frac{Q^2}{\Lambda^2}=\left ({\lambda \over {24 \pi^2}}\right )^{1/2} \frac{1}{\theta\Lambda^2}<<1 
\ee
with small $\lambda$ and large, finite $\theta\Lambda^2$;

2) $G(x,y)$ translational invariant with $G(\vec p)$ as in  Eq. (\ref{gi}) with $M^2(\vec p)$ given by 
\be\label{mdef1}
M^2(\vec p)= M^2_o 
+{ {A^2 \lambda} \over {12}}\left \lbrack 1 + {1\over 2}\cos \left (\vec p \cdot ( \vec Q  \times \vec \theta)
\right ) \right \rbrack
+{ {\lambda} \over {6\pi^2 }}\frac{1}{|\vec p  \times \vec \theta|^2}
~~~~~~~~~~~~~~~~~for~ p^2\leq Q^2
\ee 
\be\label{mdef2}
M^2(\vec p)= M^2_o ~~~~~~~~~~~~~~~~~~~~~~~~~~~~~~~~~~~~~~~~~~~~~~~~~~~~~~~~~~~~~~~~~~~~for~ p^2\geq Q^2
\ee

This choice is motivated by  the selfconsistent asymptotic solution of Eq. (\ref{trentaquattro}) (which is displayed in Eq. (\ref{masy}))  and, above all, by
the previous observation that  the transition is mainly driven by $Q^2\neq 0$ and depends weakly on the 
other details of our ansatz for $M^2(\vec p)$ in the small $p$ region.
 
By means of  Eqs. (\ref{venticinque}) and (\ref{qcond},\ref{mdef1},\ref{mdef2}), we computed, in cutoff units, 
\be
\Delta E(A,M_o^2,\theta)=E(A,M_o^2,\theta)-E(0,0,0)
\ee
for $\lambda$ and $\theta\Lambda^2$ fixed and studied the occurrence of the phase transition by changing the mass parameter $\mu^2$.

For $\mu^2=0$ there is no spontaneous symmetry breaking. Figure 2 shows $\Delta E$ for $\mu^2=0$, $\lambda=10^{-2}$ 
and $\theta\Lambda^2=100$, for different values of $M^2_0$ in the range between $10^{-2}$ and $10^{-6}$.

For  values of $\mu^2$ below a negative threshold, we observe spontaneous symmetry breaking. In Figure 3 different plots of  $\Delta E$ 
are reported for $\mu^2=-5~10^{-4}$ and again  $M^2_0$ in the range between $10^{-2}$ and $10^{-6}$. $\lambda$ and $\theta\Lambda^2$
are the same as in the previous case and the absolute minimum is for $A\sim 0.55$.

To follow the behavior of $\Delta E$ as a function of $\mu^2$, in Figure 4 we give $\Delta E$ for fixed $M^2_o=10^{-2}$ and different 
values of $\mu^2$. The absolute minimum goes slowly to zero around $\mu^2\sim -2~10^{-4}$. 
The qualitative picture does not strongly depend on the particular values taken for $\lambda$ and $\theta\Lambda^2$ as long as we stay in the 
low $Q^2$ region where the whole approach is selfconsistent.

\section  { CONCLUSIONS}  

The variational Raileigh-Ritz approximation to the CJT Effective Action  shows that for large $\theta\Lambda^2$ i.e. small $Q^2$
the transition to a broken stripe phase occurs. The mass generated by the gap equation depends on the mutual direction among $\vec \theta$,
$\vec Q$ and the momentum vector $\vec p$. This phenomenon occurs also in noncommutative electrodynamics \cite{jacpi,iorio}
where for the electromagnetic waves the modified dispersion relation 
\be 
\frac{\omega}{c}=|\vec k| \left( 1-\vec b_T \cdot \vec \theta_T \right )
\ee
depends on the angles among the wave vector $\vec k$ and the transverse components ( with respect to $\vec k$) of the background magnetic
field  $\vec b$  and of the vector $\vec \theta$.

In our approximated numerical analysis we checked that the occurrence of spontaneous symmetry breaking is weakly related 
to the ansatz made on $M^2(\vec p)$ in the small $p$ region. The transition essentially depends on $Q^2\neq 0$ and this is the 
{\it a posteriori} main motivation to believe that also a translational invariant approximation for the propagator can reproduce
the qualitative features of the transition  also of  the noncommutative  theory.
On the other hand, due to our coarse grained ansatz on the propagator,  
 we are not able to make a precise statement on the order of the phase transition which, in turn,  depends 
on the dynamical details of the theory. 

Finally we considered a simplest periodic structure for the background field in Eq. (\ref{trentatre}) because no qualitative changes
 are expected for more
complicated superpositions as suggested in \cite{gubser}. 

\vspace{0.8cm}

{\bf Acknowledgements}
We are indebted with S.-Y. Pi for constant advice and for many remarks about the manuscript.
We thank Roman Jackiw for many fruitful suggestions. 
We are also grateful to M. Consoli and L. Griguolo for helpful discussions.
This work, started during a visit of the authors to the Center for Theoretical Physics, MIT, 
is supported in part by funds provided by the U.S.
Department of Energy (D.O.E.) under cooperative research agreement
DF-FC02-94ER40818.

\begin{figure}
\epsfig{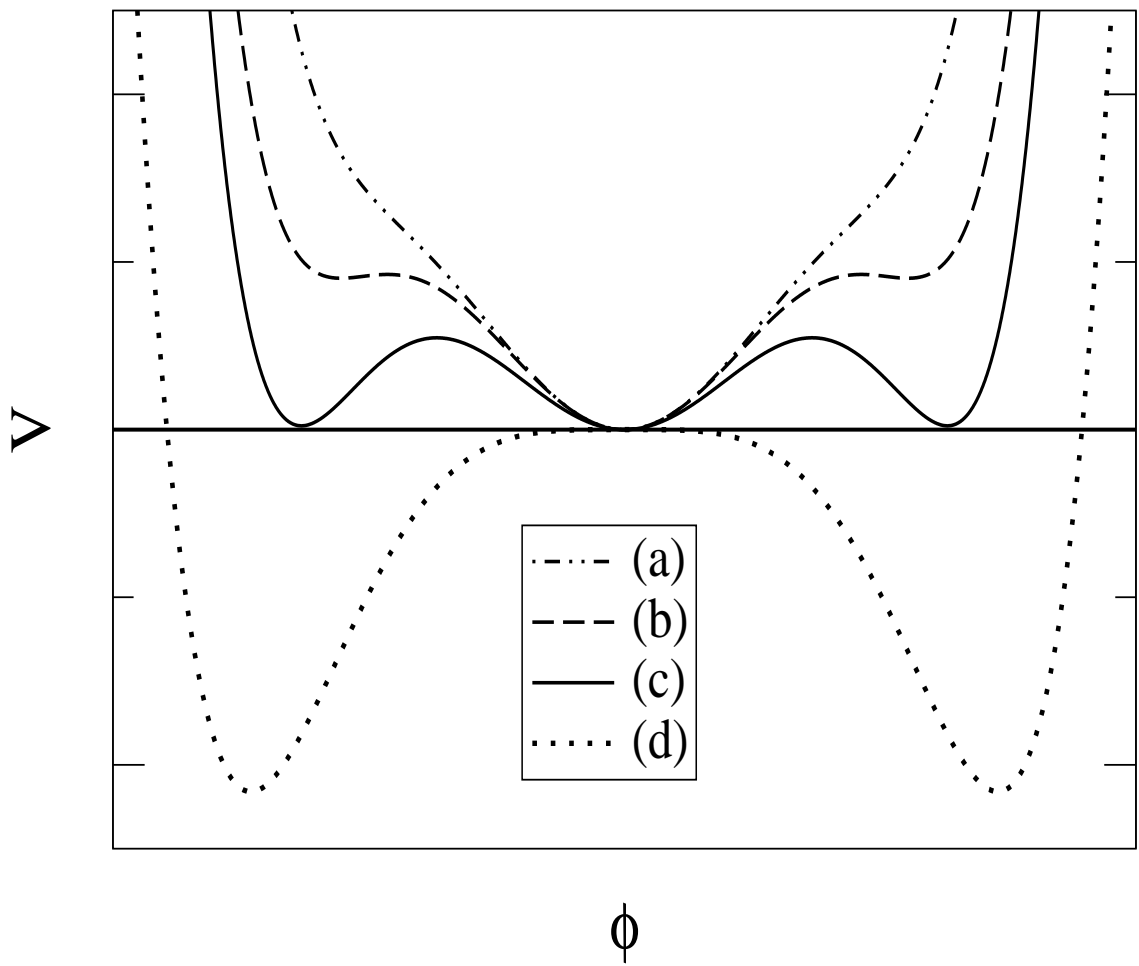}
\caption{The Effective Potential of the scalar commutative
theory in the HF approximation for various values of the parameter $\mu$ (see text for details).
}
\end{figure}

\eject

\begin{figure}
\epsfig{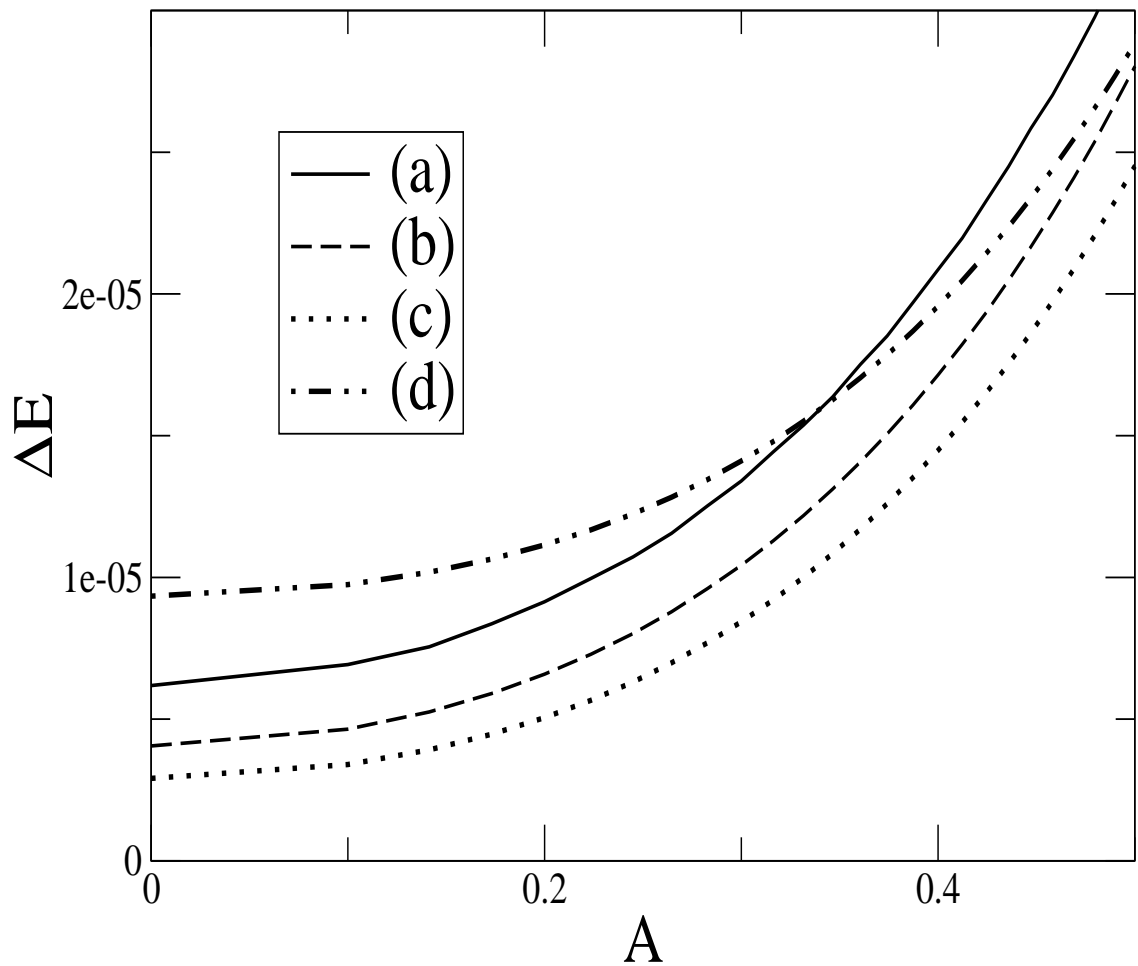}
\caption{$\Delta E$  vs. $A$  with $\lambda=10^{-2}$, $\theta=100$, $\mu^2=0$
and $M^2_o=10^{-6}$ (a), $M^2_o=10^{-3}$ (b), $M^2_o=8 ~10^{-3}$ (c), $M^2_o=4 ~ 10^{-2}$ (d).
}
\end{figure}

\eject

\begin{figure}
\epsfig{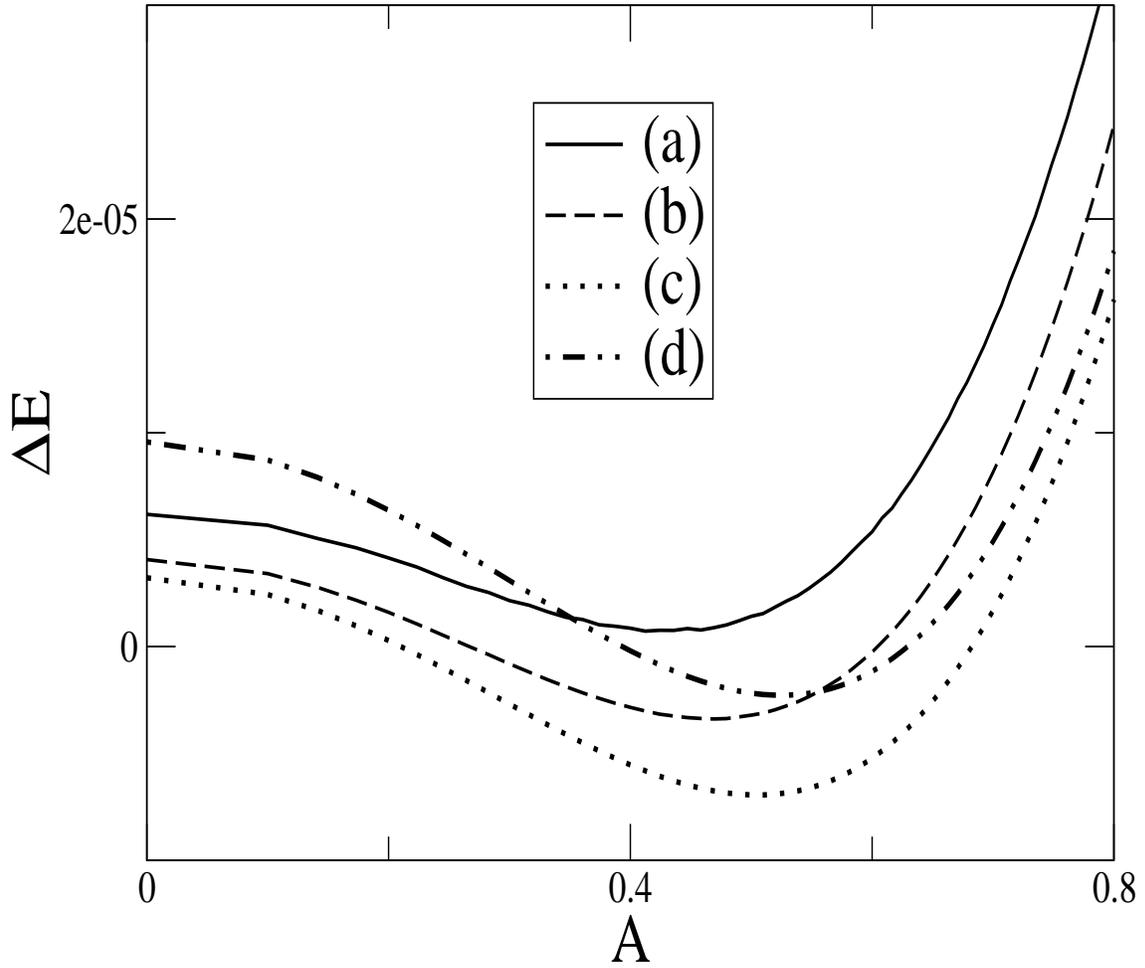}
\caption{$\Delta E$  vs. $A$  with $\lambda=10^{-2}$, $\theta=100$, $\mu^2=-5 ~10^{-4}$
and $M^2_o=10^{-6}$ (a), $M^2_o=10^{-3}$ (b), $M^2_o=1.2 ~10^{-2}$ (c), $M^2_o=4 ~10^{-2}$ (d).
}
\end{figure}

\eject

\begin{figure}
\epsfig{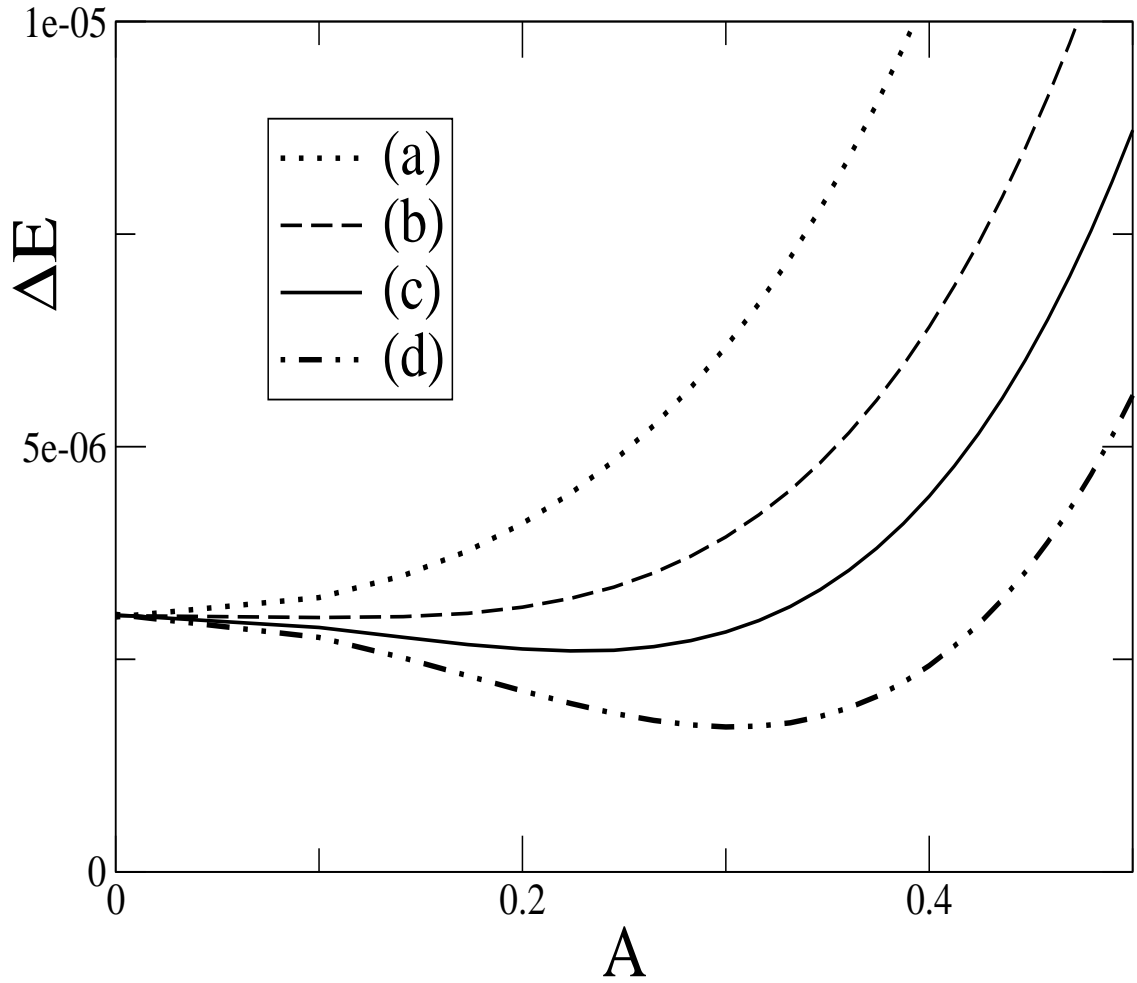}
\caption{$\Delta E$  vs. $A$  with $\lambda=10^{-2}$, $\theta=100$, $M_0^2=10^{-2}$
and $\mu^2=-10^{-4}$ (a), $\mu^2=-2 ~10^{-4}$ (b), $\mu^2=-2.5~ 10^{-4}$ (c), $\mu^2=-3~ 10^{-4}$ (d).
}
\end{figure}
   

\begin{thebibliography}{18}
\bibitem{connes}
A. Connes, M.R. Douglas, A. Schwarz, JHEP {\bf 02} (1998) 003.
\bibitem{witten}
N. Seiberg and E. Witten, JHEP {\bf 09} (1999) 032.
\bibitem{camp}
B.A. Campbell and A. Kaminsky , Nucl Phys. {\bf B581} (2000) 240.
\bibitem{gubser}
S.S. Gubser and  S.L. Sondhi, Nucl. Phys. {\bf B605}  (2001) 395.
\bibitem{chen}
Guang-Hong Chen, Yong-Shi Wu, Nucl. Phys.  {\bf B622} (2002) 189. 
\bibitem{rivelles}
``Spontaneous symmetry breaking in noncommutative field theory'' 
H.O. Girotti, M. Gomes, A.Yu. Petrov, V.O. Rivelles, A.J. da Silva, Preprint Jul 2002.
e-Print Archive: hep-th/0207220.
\bibitem{bieten} ``Simulating noncommutative field theory'', 
W. Bietenholz, F. Hofheinz, J. Nishimura,
Preprint : HU-EP-02-35, Sep 2002.  e-Print Archive: hep-lat/0209021. 
\bibitem{catterall}
J. Ambjorn, S. Catterall, Phys. Lett. {\bf B549} (2002) 253;\\
``Noncommutative field theories beyond perturbation theory''. 
W. Bietenholz, F. Hofheinz, J. Nishimura,
Preprint : HU-EP-02-63, Dec 2002.  e-Print Archive: hep-th/0212258.
\bibitem{cjt}
J. M. Cornwall, R. Jackiw and E. Tomboulis, Phys Rev. {\bf D 10} (1974) 2428.
\bibitem{minwalla}
S. Minwalla, M. Van Raamsdonk an N. Seiberg, JHEP {\bf 0002} (2000) 020.
\bibitem{mvan}
M. Van Raamsdonk an N. Seiberg, JHEP {\bf 0003} (2000) 035.
\bibitem{chepelev} I. Chepelev and R. Roiban, JHEP {\bf 0103} (2001) 001.
\bibitem{griguolo} L. Griguolo, M. Pietroni, JHEP {\bf 0105} (2001) 032.
\bibitem {doplicher}
D. Bahns, S. Doplicher, K. Fredenhagen and G. Piacitelli, Phys. Lett {\bf B533} (2002) 178.
\bibitem{ruiz}
F. Ruiz Ruiz, Nucl. Phys.  {\bf B637}  (2002) 143.
\bibitem{pi1} 
So Young Pi and M. Samiullah, Phys Rev. {\bf D 36} (1987) 3128. 
\bibitem{noi}
V. Branchina,P.Castorina,M.Consoli and D.Zappal\`a, Phys Rev. {\bf D 42} (1990) 3587.
\bibitem{braz} S.A. Brazovskii, Zh. Eksp. Teor. Fiz.  {\bf 68} (1975)175.
\bibitem{nota}
It should be noted that there is a solution of Eq. (\ref{sedici}), $M^2>1$, both for vanishing and
nonvanishing $\mu^2$, which means a mass larger than the ultraviolet cutoff. Since we require
our parameters to be smaller than the cutoff, we shall not retain this solution in our analysis.
\bibitem{polch}
J Polchinski,  Nucl. Phys. {\bf B231}  (1984) 269.
\bibitem{douglas}
M.Douglas and N.Nekrasov, Rev.  Mod. Phys. {\bf 73} (2001) 977. 
\bibitem{jacpi} Z.Guralnik, R. Jackiw and S. Y. Pi, Phys. Lett. {\bf 517} (2001) 450.
\bibitem{iorio} ``Noncommutative synchrotron'',  P. Castorina, A. Iorio, D. Zappal\`a,
Preprint : MIT-CTP-3336, Dec 2002.  e-Print Archive: hep-th/0212238. 
\end{thebibliography}
\end{document}